\documentclass[twocolumn,showpacs,preprintnumbers,amsmath,amssymb]{revtex4}

\usepackage{etex}
\usepackage{graphicx}
\usepackage{dcolumn}
\usepackage{bm}
\usepackage{dsfont}
\usepackage{amsthm,amscd,amsbsy,array}
\usepackage{color}

\newcommand{\half}{{\scriptstyle{\frac{1}{2}}}}
\def\2{{\half}}
\newcommand{\const}{\mathop{\rm const}\nolimits}

\def\bR{{\mathds{R}}}

\def\bp{{\bf p}}

\def\bu{{\bm{u}}}

\def\br{{\bm{r}}}
\def\bQ{{\bm{Q}}}
\def\bP{{\bm{P}}}
\def\bq{{\bm{q}}}
\def\bp{{\bm{p}}}
\def\beq{\begin{equation}}
\def\eeq{\end{equation}}
\def\beqa{\begin{eqnarray}}
\def\eeqa{\end{eqnarray}}

\def\barray{\left(\begin{array}}
\def\earray{\end{array}\right)}
\def\barraynb{\begin{array}}
\def\earraynb{\end{array}}

\def\smallover#1/#2{\hbox{$\textstyle\frac{#1}{#2}$}} %

\newcommand{\Sp}{\mathrm{Sp}}
\newcommand{\cJ}{\mathcal{J}}

\newcommand{\sfp}{\mathsf{p}}
\newcommand{\cP}{\mathcal{P}}

\newcommand{\sfs}{\mathsf{s}}

\newcommand{\iin}{\mathrm{in}}
\newcommand{\out}{\mathrm{out}}

\begin{document}

\preprint{
arXiv:1202.2430v3
}

\title{Transverse Shifts in Paraxial Spinoptics\footnote{Dedicated to the memory of J.-M. Souriau, deceased on March 15 2012.}}

\author{
C. Duval$^{1}$\footnote{Aix-Marseille Univ, CNRS UMR-7332,  Univ Sud Toulon-Var,
13288 Marseille Cedex 9, France. mailto:duval@cpt.univ-mrs.fr},
P. A. Horvathy$^{2,3}$\footnote{mailto:horvathy@lmpt.univ-tours.fr},
P. M. Zhang$^{2}$\footnote{mailto:zhpm@impcas.ac.cn}
}

\affiliation{
$^1$Centre de Physique Th\'eorique, 
Marseille, France
\\
$^2$Institute of Modern Physics, Chinese Academy of Sciences,
Lanzhou, China 
\\
$^3$Laboratoire de Math\'ematiques et de Physique
Th\'eorique,
Universit\'e de Tours, 
France}

\date{\today}

\begin{abstract}
The paraxial approximation of a classical
spinning photon is shown to yield an ``exotic particle'' in the  plane transverse to the propagation. The previously proposed and observed  \emph{position shift} between media with different refractive indices is  modified 
when the interface is curved, and there
 also appears
a novel, \emph{momentum} [direction] shift. The laws of thin lenses are modified accordingly. 
\end{abstract}


\maketitle


Fermat's light is simply a line in space.
Real light has also spin, however. 
A first consequence is that the propagation in space induces a change
of the polarization \cite{Rytov,Chiao}.
More recently, it has been argued that the change of the polarization should be fed back at  the semiclassical dynamics of light \cite{OptiHall}. Then the most dramatic consequence is that the outgoing ray is slightly shifted in the transverse direction with respect to the incoming light ray \cite{OptiHall, DHSpinOptics}. This recently  observed \cite{HostenKwiat,OptiHallEx}  spin-Hall-type effect occurs when the interface is a plane; then the proof relies on the symmetries of the geometrical setting, namely momentum and angular momentum conservation.
See, e.g., \cite{BA} for reviews.

In this Letter we point out that at a curved interface with no particular symmetry, as that of a lens, for example, an additional effect arises~: the Hall shift is modified  and also the \emph{direction} of the light ray is  deflected due to the curvature.
 
We demonstrate our statement by generalizing the
paraxial approximation from scalar to spin optics.


We first consider \emph{spinless} light rays propagating approximately along the $z$ axis in the forward direction.
In matrix optics \cite{Gerrard,GSt} a light ray is labeled
 by the point $\bQ$ where it hits some
given reference plane, and by its direction
at that point, 
given by a vector $\bu$. It is more convenient
to use instead the ``momentum'' $\bP=n\bu$, where $n$ is the the refractive index, 
 assumed to be locally constant and discontinuous 
 on the interfaces. The $2$-vectors $\bQ$ and $\bP$ label a point in phase space, i.e., label a light ray in the paraxial approximation.

We choose incoming and outgoing reference planes at $z=z_\iin{}$ and $z=z_\out{}$ such that 
our optical device lies between the two.
Light propagation through our device is described, classically, by a \emph{canonical transformation} (or \emph{symplectic scattering}) between  ``in'' and ``out'' states~\cite{SSD}.
It is hence a classical counterpart of the
S-matrix in quantum mechanics.
In the linear (or paraxial) approximation our optical instrument is characterized
therefore by a  $4\times4$ matrix $M$ such that $M^t\cJ{}M=\cJ$, where
$
\cJ=
\barray{cc}
0&-1\\1 &0
\earray
$ 
is the canonical symplectic matrix of $4$-dimensional phase space, given by Eq. (\ref{omegabis}) below.
The ingoing and outgoing rays are hence related by \cite{Gerrard,GSt} 
\begin{equation}
\barray{c}
\bQ_\out{}\\
\bP_\out{}
\earray
=
M
\barray{c}
\bQ_\iin{}\\
\bP_\iin{}
\earray.
\label{M}
\end{equation}
An  \emph{optical interface}, 
for example, is given by a refraction matrix
\begin{equation}
L=
\barray{cc}
1&0\cr
-\cP&1
\earray,
\label{L}
\end{equation}
where the symmetric $2\times2$ matrix~$\cP$ is the \emph{power} of the device. For a plane interface, for example, $\cP=0$, so $M$ is the unit matrix.  Thus $\bP_\out{}=n_\out{}\bu_\out{}=
\bP_\iin{}=n_\iin{}\bu_\iin{}$ which is the linearized form of Snel's law of refraction. The latter is hence incorporated into the formalism.

For a cylindrical-symmetric thin lens $\cP$  is a scalar 
determined by the refractive indices, $n_\out{}$ and $n_\iin{}$ of the lens and the surrounding optical medium, respectively, and 
the (signed) curvatures of the lens.

Translations matrices are in turn 
of the form
$T=\barray{cc}1&d\cr0&1\earray$,
where the scalar $d>0$ represents the {optical length} along which light travels freely in the optical medium of constant refractive index $n$ \cite{Gerrard,GSt}.
The important  result of \cite{GSt} is that any optical instrument, i.e., any element of the symplectic group $\Sp(4,\bR)$ can be decomposed as a product of matrices of type $L$ and~$T$, the building blocks of linear optics.

In view of  Eq. (\ref{M}) with $M=L$, refraction is described by
\begin{equation}
\barray{c}
\bQ_\out{}\\
\bP_\out{}
\earray
=
\left(
\begin{array}{l}
\bQ_\iin{}\\
\bP_\iin{}-\cP \bQ_\iin{}
\end{array}
\right),
\label{Lbis}
\end{equation}
so that $\bQ_\out{}=\bQ_\iin{}$, confirming that spinless light rays fall on and get out of the lens at the same point. The equation 
\beq
\bP_\out{}=\bP_\iin{}-\cP \bQ_\iin{}
\label{0momshift}
\eeq
is in turn an expression of the properties of the optical device, 
and follows from [the linearized] Snel law of refraction \cite{Gerrard,GSt}.


Likewise, for a translation, i.e., if $M=T$, we get
$\bP_\out{}=\bP_\iin{}$, and  $\bQ_\out{}=\bQ_\iin{}+d\,\bP_\iin{}$, which is clearly the new location of the photon, away, at the distance $d$.

\goodbreak


{As first proposed by Souriau} \cite{SSD},  semiclassical
free spinning photon is described 
 by the $2$-form 
\begin{equation}
\sigma=d(\sfp\,\bu\cdot{}d\br)-\sfs\,\Omega\,,
\label{sigma}
\end{equation}
where $\sfp=\hbar/\lambdabar$ (\emph{color}), 
and $\sfs=\chi\hbar$ (\emph{spin}) 
are Euclidean invariants of the model; here $\lambdabar$ is the reduced wavelength, and $\chi=\pm1$ the helicity
\cite{SSD,DHSpinOptics}. In Eq. (\ref{sigma}), 
the $3$-vector~$\br$ is an arbitrary point on the light
ray, whose direction is the unit $3$-vector $\bu$; also $\Omega=\half\epsilon_{ijk}u_i\cdot du_j\wedge du_k$ is the area element of the $2$-sphere. 

Such a description is clearly redundant, as two vectors $\br$ and $\br'$  represent the same ray whenever $\br-\br'$ is proportional to $\bu$. 
A light ray  corresponds rather to  a null curve $\gamma$ of $\sigma$, i.e., such that $\sigma_{\alpha\beta}\,\dot{\gamma}^\alpha=0$.  
In the free case, these  are indeed straight lines, pointing in the direction of $\bu$. 

We just mention here that the 2-form $\sigma$ in (\ref{sigma}) is related to a generalized variational calculus in a suitably extended space; the first term corresponds to the usual Fermat term of a ``spinless photon'' while the second, ``Berry'' term, represents the spin  \cite{DHSpinOptics}.

The ``gravitational'' coupling of our photon to the Fermat metric $ds^2=n^2(\br)(dx^2+dy^2+dz^2)$ yields \cite{DHSpinOptics}  Papapetrou-type equations for spinning light rays in an inhomogeneous isotropic medium characterized by a refractive index $n$. The latter duly reduce to those of \cite{OptiHall} in the special case of circularly polarized light coupled to a slowly varying refractive index.

At our (semi-)classical level, 
the paraxial approximation amounts to  converting the $z$ coordinate into time by a trick reminding one to taking the non-relativistic limit \cite{JaNa}. 
Let us hence introduce 
$u_x={v_x}/{c}$
and
$u_y={v_y}/{c}$,
where $c>0$ has the dimension of velocity. 
Then 
$u_z=
1-(v_x^2+v_y^2)/2c^2+\cdots$,
where the ellipses denote higher-order terms in $c^{-2}$.
Inserting into (\ref{sigma}) and defining
$ 
m=\sfp/{c}
$ 
we find, 
\begin{equation}
\sigma
=m(dv_x\wedge{}dx + dv_y\wedge{}dy)-dE\wedge{}dt+\kappa\,dv_x\wedge{}dv_y+\cdots
\label{sigmaBis}
\end{equation}
where
$
t={z}/{c}
$
and
$
E=({m}/{2})(v_x^2+v_y^2)-E_0 +\cdots
$
with
$
E_0=mc^2
$
are a time coordinate and non-relativistic energy, respectively,
and where 
$
\kappa=-\chi{\hbar}/{c^2}.
$
In (\ref{sigmaBis})  we recognize the $2$-form
of the  ``exotic''  planar model 
constructed by Souriau's method \cite{SSD}, starting with the
two-fold extension of the planar Galilei group  
\cite{DHexo,JaNa}~:
\emph{par\-axial approximation converts  
free spinoptics into ``exotic'' classical mechanics in the transverse $(x,y)$-plane.}
Ordinary geometrical optics is recovered in the limit $\hbar=0$.

%

To describe light in an optical medium it is
enough to replace the color, $\sfp=\hbar/\lambdabar$, by
$n\sfp$ according to the Fermat prescription. 

Let us start with the case $n=\const$.
Putting $t=0$, in  (\ref{sigmaBis}), we get a bona fide symplectic $2$ form,
$\omega
$, on the phase space. 
After a suitable rescaling, 
{$\omega\to(\lambdabar/\hbar)\omega$}, 
introducing 
$p_x=n u_x, p_y=n u_y$,
we end up with
\begin{equation}
\omega
\approx 
dp_x\wedge{}dx+dp_y\wedge{}dy+\kappa\,dp_x\wedge{}dp_y,
\quad
\kappa=-\frac{\chi\lambdabar}{n^2
}\,.
\label{homega}
\end{equation}
Paraxial approximation of  spinoptics is therefore governed by the \emph{twisted} (\emph{exotic}) symplectic form (\ref{homega}); the Hamiltonian $H=E/\sfp{}c$ is 
$
H\approx(p_x^2+p_y^2)/{2n}-n.
$

A tedious calculation, performed along the lines of \cite{DHSpinOptics}, shows that the preceding formulas for $\omega$ and $H$ still hold for a slowly varying refractive index, $n$. Below, we shall only apply our formalism to the case of a constant refractive index, and use symplectic scattering across some sharp interfaces, where it jumps. 

Consistently with Darboux's theorem, the
 twisted symplectic
structure (\ref{homega}) can be brought into canonical form. Put indeed
\begin{equation}
\barray{c}
\bQ\cr
\bP
\earray
=
\barray{cc}
1&\half\kappa J\cr 0&1
\earray
\barray{c}
\bq\cr 
\bp
\earray,
\label{1+Z}
\end{equation}
where where $\bq=(x,y)$ and $\bp=(p_x,p_y)$, and ditto for $\bQ$ and $\bP$.
Here, $J=(\epsilon_{ij})$ is a rotation by $90$ degrees clockwise in the plane. The transformation (\ref{1+Z}) brings the symplectic matrix  (\ref{homega}) into  canonical form,
\begin{equation}
\omega\approx dP_x\wedge{}dX+dP_y\wedge{}dY.
\label{omegabis}
\end{equation}


Now our fundamental assumption is that for paraxial optics whose
symplectic matrix is canonical i.e., for
the coordinates $(\bQ,\bP)$,
 the previously recorded laws of linear optics should apply.
 Then the results should simply be translated to our \emph{physical} coordinate system
$(\bq,\bp)$. 

The scattering of light by the (curved) interface between two regions with refractive indices $n_\iin{}$, and $n_\out{}$, i.e., with power $\cP_\iin{}$, reads, hence,
\begin{equation}
\barray{c}\bq_\out{}\\ \bp_\out{}\earray
=
\left(\begin{array}{l}
\bq_\iin{}+\half{}J\left(\kappa_\iin{}\bp_\iin{}-\kappa_\out{}\bp_\out{}\right)\\[4pt]
\bp_\iin{}-\cP_\iin{}\left(\bq_\iin{}+\half\kappa_\iin{}{}J\bp_\iin{}\right)
\end{array}
\right).
\label{S}
\end{equation}
This entails that the impact location of the ray undergoes a  spin-Hall shift across the interface, viz., 
\begin{equation}
\bq_\out{}-\bq_\iin{}=\frac{1}{2}J
(\kappa_\iin{}\bp_\iin{}-\kappa_\out{}\bp_\out{}).
\label{opticalHall}
\end{equation}

From Eq. (\ref{S}), we deduce, furthermore, the additional \emph{momentum shift} $\Delta{\bp}=\bp_\out{}-\bp_\iin{}+\cP_\iin\bq_\iin$, namely
\begin{equation}
\Delta{\bp}=-\half\kappa_\iin{}{}\cP_\iin{}J\bp_\iin{}
\label{momshift}
\end{equation}
that yields a modification of  (\ref{0momshift}).
But the momentum is proportional to the direction, $\bp=n\bu$, so this  amounts in fact to 
\emph{deflecting the direction} of the
outgoing ray. 

\goodbreak

As to the position shift, $\Delta{\bq}=\bq_\out{}-\bq_\iin{}$, it now reads
\begin{equation}
\Delta{\bq}=\half J\left((\kappa_\iin-\kappa_\out)\bp_\iin+\kappa_\out\cP_\iin\bq_\iin\right)+\cdots
\label{deltaq}
\end{equation}
up to a term $\sim\lambdabar^2$.

Restricting our considerations to, e.g.,  cylindrically symmetric interfaces for which $\cP_\iin{}=1/f_\iin{}$ is a scalar matrix and 
putting, e.g., $\bp_\iin{}=(n_\iin{}\theta_\iin{},0)$ where $\theta_\iin{}\ll1$ is the angle of incidence, we readily get from Eq. (\ref{deltaq})
\begin{eqnarray}
\label{Deltax}
\Delta{x}
&\cong&
-\frac{\chi\lambdabar}{2n_\out^2}\,\frac{y_\iin}{f_\iin},
\\[4pt]
\Delta{y}
&\cong&
\chi\lambdabar\left[\frac{n_\out{}^2-n_\iin{}^2}{2\,n_\iin{}n_\out{}^2}\right]\theta_\iin{}+\frac{\chi\lambdabar}{2n_\out^2}\,\frac{x_\iin}{f_\iin}\,,
\label{Deltay}
\end{eqnarray}
which, for $f_\iin\to+\infty$, reduces to
 the linearized \emph{optical Hall shift}  \cite{OptiHall,DHSpinOptics}, viz.,
\beq
\Delta{x}=0,
\qquad
\Delta{y}
=
\chi\lambdabar\frac{\cos\theta_\out{}-\cos\theta_\iin{}}{n_\iin{}\sin\theta_\iin{}}\,.
\label{Hallshift}
\eeq
The extra terms in (\ref{Deltax}) and
(\ref{Deltay}) arise from curvature,
and disappear for $\bq_\iin=0$.
From (\ref{momshift}) we infer that  
$\Delta{\bp}$ is perpendicular to
the in\-coming momentum,
\begin{equation}
\Delta{p_x}
=0,
\qquad
\label{Deltapxy}
\Delta{p_y}
=
-\frac{\chi\lambdabar}{2\,n_\iin{} f_\iin{}}\,\theta_\iin{}.
\end{equation}

Let us apply our results to the example of a  thin lens. 
Denoting by $\cP_\iin{}$, and $\cP_\out{}$ the  powers of the ``in'' and ``out'' interfaces  with $z_\iin\cong z_{\out}$, we readily get $\cP=\cP_\iin{}+\cP_\out{}$.
Then, the preceding formula (\ref{S}), applied successively, 
 leads  to  the following position and momentum shifts, 
\begin{equation}
\Delta\bq
\cong
+\half\kappa_\iin J\cP\bq_\iin,
\qquad
\Delta\bp
=
-\half\kappa_\iin \cP J\bp_\iin.
\label{ShiftsThinLens}
\end{equation}


In conclusion, our
 main result  is that when polarized light is refracted by
an optical device, the rays suffer, in addition to the
already confirmed optical Hall shift \cite{HostenKwiat,OptiHallEx}, an extra positional shift due to curvature, as well as a deflection of their \emph{direction}.
Both effects are of the order of the wavelength. 

The optical Hall shift, (\ref{Hallshift}), is usually derived \cite{OptiHall,DHSpinOptics} from the conservation of the angular momentum, \emph{assuming} Snel's law -- derived in turn from \emph{linear momentum conservation} \cite{DHSpinOptics}. Angular momentum is also present for  axially symmetric lenses; it is the \emph{translational symmetry} along the
interface which is broken.

The momentum effect only arises for optical devices with non-trivial power, $\cP\ne0$,
and vanishes for a planar interface. 

Intuitively, due to
the position displacement, the incoming ray goes out at a  position which is slightly different from where it entered; but this implies a different incidence angle, since the tangent plane has moved, too. 

{ It is worth mentioning that our effect here
is different from the momentum shift predicted in \cite{BPRE}, since the latter is refraction-related,
and is rather reminiscent of the focal shift
studied in \cite{BPRA10,Zeldo}.}

We mention for completeness that for transla\-tions, applied to spinoptics, we merely get $\bq_\out{}=\bq_\iin{}+d\,\bp_\iin$, and $\bp_\out{}=\bp_\iin{}$ as expected.

In its common formulation, paraxial optics means converting Maxwell's equations 
into the form of a non-relativistic Schr\"odinger equation with the $z$ coordinate
playing the r\^ole of time. In this Letter we use, instead, a
(semi-)\emph{classical} mechanical model, whose ``quantization''
would allow us to recover the Schr\"odinger description.

\begin{acknowledgments} 
We would like to thank K. Bliokh for enlightening correspondence, and G. Gibbons for directing our attention to the subject.
P.A.H is indebted to the \emph{Institute of Modern Physics} of the Lanzhou branch of
the Chinese Academy of Sciences for hospitality.
This work was partially supported by the National Natural Science Foundation of 
China (Grant No. 11035006) and by the Chinese Academy of Sciences Visiting 
Professorship for Senior International Scientists (Grant No. 2010TIJ06). 
\end{acknowledgments}




\begin{thebibliography}{99}

\bibitem{Rytov}
 S. M. Rytov, 
 Dokl. Akad. Nauk. SSSR {\bf 18} (1938) 263;
V. V. Vladimirskii,  Dokl. Akad. Nauk. SSSR 
{\bf 31} (1941) 222.

\bibitem{Chiao}
R. Chiao and Y-S Wu, 
Phys. Rev. Lett. {\bf 57} (1986) 933;
A. Tomita and R. Chiao,
 Phys. Rev. Lett. {\bf 57} (1986) 937.
  
\bibitem{OptiHall}   
K. Yu. Bliokh and Yu. P. Bliokh,
 JETP Lett. {\bf 79} (2004) 519;
 Phys. Lett.  {\bf A333}  (2004) 181, \texttt{physics/0402110};
M. Onoda, S. Murakami and N. Nagaosa,
Phys. Rev. Lett. {\bf 93}  (2004), 083901.

\bibitem{DHSpinOptics}
C. Duval, Z. Horv\'ath, P. A. Horv\'athy~:
Phys. Rev. {\bf D74}, 021701(R) (2006), 
 Journ. Geom. Phys. {\bf 57},  925 (2007).

\bibitem{HostenKwiat} 
O. Hosten and P. Kwiat,
Science {\bf 319}, 787 (2008).

\bibitem{OptiHallEx}
 K.~Y~Bliokh., A. Niv, V.~Kleiner and  E.~Hasman,
Nature Photonics {\bf 2} (2008), 748,
\texttt{arXiv:0810.2136}.

\bibitem{BA}
K. Y. Bliokh and A. Aiello,
``Goos-H\"anchen and Imbert-Fedorov beam shifts at a plane dielectric interface'',
Laser Physics (to be published);
K. Y. Bliokh, A. Aiello, and M. A. Alonso,
``Spin-orbit interactions of light in isotropic media''. Book chapter (to be published).
   
\bibitem{SSD}
J.-M.~Souriau,
\textsl{Structure des syst\`emes dynamiques}, Dunod (1970). 
\textsl{Structure of Dynamical Systems. A Symplectic View of Physics},
Birkh\"auser, Boston (1997).

\bibitem{Gerrard}
A. Gerrard, J. M. Birch,
\textsl{Introduction to matrix methods in optics}. Wiley-Interscience Publication 
London - New York - Sidney - Toronto (1975).

\bibitem{GSt}
V. Guillemin and S. Sternberg,
\textsl{Symplectic techniques in physics},
Cambridge UP (1984).
  
\bibitem{DHexo}
C.~Duval and  P.~A.~Horv\'athy, 
Phys. Lett. {\bf B 479} (2000) 284; 
 Journ. Phys. {\bf A 34}  (2001) 10097,

\bibitem{JaNa}
 R.~Jackiw and  V.~P.~Nair, 
Phys. Lett. {\bf B 480} (2000) 237,
 C. Duval and  P. A. Horv\'athy,
Phys. Lett. {\bf B 547} (2002) 306. 

\bibitem{BPRE}
K. Yu. Bliokh and Yu. P. Bliokh,
Phys. Rev. E 75, 066609 (2007)

\bibitem{BPRA10}
K. Y. Bliokh, M. A. Alonso, E. A. Ostrovskaya, and A. Aiello,
Phys. Rev. {\bf A 82}, 063825 (2010)

\bibitem{Zeldo}
B. Ya. Zeldovich, N.D. Kundikova, 
Pisma Zh. Eksp. Teor. Fiz. {\bf 59} (1994) 737. 
 
\end{thebibliography}
\end{document}